\documentclass[twocolumn,preprintnumbers,amsmath,amssymb]{revtex4}

\usepackage{graphicx}
\usepackage{dcolumn}
\usepackage{bm}

\usepackage{color}
\usepackage{placeins}

\usepackage[utf8]{inputenc}
\usepackage[T2A]{fontenc}

\usepackage[dvipsnames]{xcolor}

\usepackage[normalem]{ulem}
\usepackage[colorlinks=true,citecolor=blue,linkcolor=blue, urlcolor=cyan]{hyperref} 

\begin{document}

\title{
Nonreciprocal  magnetic-field-induced second harmonic generation of exciton polaritons in ZnSe
}

\author{J. Mund$^1$, D. R. Yakovlev$^{1,2}$, A. Farenbruch$^1$, N. V. Siverin$^1$,  M. A. Semina$^2$, M. M. Glazov$^2$, E. L. Ivchenko$^2$, and M. Bayer$^{1,3}$}
\affiliation{$^{1}$ Experimentelle Physik 2, Technische Universit\"at Dortmund, 44227 Dortmund, Germany}
\affiliation{$^2$ Ioffe Institute, Russian Academy of Sciences, 194021 St. Petersburg, Russia}
\affiliation{$^3$ Research Center FEMS, Technische Universit\"at Dortmund, 44227 Dortmund, Germany}

\date{\today}

\begin{abstract}
We report on the optical second harmonic generation (SHG) on the 1S exciton-polariton resonance in bulk ZnSe that is subject to an external magnetic field applied perpendicular to the light wave vector $\mathbf k$ (Voigt geometry). For the symmetry allowed geometry with the $\mathbf{k}\parallel[111]$ crystal axes, the nonreciprocal dependence of the SHG intensity on the magnetic field direction is found. It is explained by an interference of the crystallographic and magnetic-field-induced SHG signals. Relative phases of these signals are evaluated from the rotational anisotropy diagrams. Phenomenological and microscopic models of the effect are developed. To the best of our knowledge, this is the first experimental observation of the nonreciprocal SHG in semiconductor crystals, and the first one for exciton-polaritons. 
\end{abstract}

\maketitle


\section{Introduction}

In linear optics, reciprocity is kept in absence of magnetic field~\cite{Potton04,Caloz18,Zvezdin_book}. That is, the experimental outcome is invariant under exchange of source and detector.  However, reciprocity is broken in magnetic field, being illustrated by the textbook example of the Faraday rotation effect \cite{Faraday46}. Since the principle of reciprocity is not, generally speaking, sustained in nonlinear optics, second harmonic generation (SHG) is a promising tool  for observing nonreciprocal effects~\cite{Shen,Boyd}. The breaking of reciprocity in nonlinear optics has gained increased interest recently. It has been demonstrated for the antiferromagnet CuB$_2$O$_4$~\cite{Mund20-CuB4O2, Toyoda20},  the antiferromagnetic bilayer CrI$_3$~\cite{Sun:2019aa}, metalic multilayers ~\cite{Vollmer:2001aa}, metalic films~\cite{Murzina:1998aa} and metalic surfces and interfaces~\cite{Kirilyuk:05,Kolmychek:2020aa}.

SHG is a coherent process which comprises two-photon excitation and one-photon emission under the energy and momentum conservation. The two-photon induced nonlinear polarization $\textbf{P}^{2\omega}$ at twice the fundamental frequency, $2\omega$, is the source of the emitted  second harmonic electromagnetic field. It is described by
\begin{equation}\label{eq.Pol_cryst}
P^{2\omega}_i=\chi^{(2)}_{ijk}E^\omega_jE^\omega_k
\end{equation}
with the susceptibility tensor components of second order, $\chi^{(2)}_{ijk}$ where $i$, $j$, $k$ are Cartesian indices and $E^\omega_{j(k)}$ are the ingoing electric fields of the photons of fundamental frequency $\omega$ \cite{Boyd}.

Several semiconductor materials have been investigated by SHG spectroscopy in recent years with a specific focus on resonantly enhanced SHG at exciton and exciton-polariton resonances, Among then are  GaAs \cite{Pavlov05, Saenger06, Warkentin18}, CdTe \cite{Saenger06}, ZnO \cite{Lafrentz13}, Cu$_2$O \cite{Mund18, Farenbruch20}, and ZnSe \cite{Mund20-ZnSe} crystals and thick layers,  ZnSe/BeTe~\cite{Mund20-ZnSeMQW} and ZnO/(Zn,Mg)O~\cite{Mund21-ZnOQW} quantum wells, as well as WSe$_2$, WS$_2$, and MoS$_2$ two-dimensional materials~\cite{Wang,Kim,Shree}. However, the nonreciprocal SHG has not been reported so far for semiconductor crystals or their heterostructures.

For this study we choose ZnSe semiconductor crystals due to its cubic noncentrosymmetric crystal lattice (point group $\bar{4}3m$ or $T_d$). Therefore, SHG is allowed in electric-dipole approximation as described by Eq.~(\ref{eq.Pol_cryst}). We observe a nonreciprocal dependence of the intensity of SHG on the magnetic field: for $\mathbf B$ and $-\mathbf B$ the SHG intensity strongly differs. We restrict the measurements to the 1S exciton resonance to exclude overlap with other exciton states, e.g., 2S or 2P. The observed optical nonreciprocity is provided by interference of crystallographic and magnetic-field-induced SHG, which is present for light propagation along the [111] crystal axis and absent for the light propagation along the [001] axis.

\section{Experimentals}

We apply a SHG spectroscopic technique where broad-band fs-laser pulses are used to excite the sample. A 1~m-monochromator in combination with a nitrogen-cooled charge-coupled device (CCD) camera is used for high spectral resolution ($30~\mu$eV) of the signal light \cite{Mund18}. ZnSe samples are installed in a cryostat with a superconducting split-coil solenoid and cooled down to a temperature of $T=5$~K. Magnetic fields up to $B=10$~T are applied in either direction along the sample $\textbf{x}$ axis. The magnetic field direction is set perpendicular to the light direction $\textbf{k}^\omega\parallel\textbf{z}$ (Voigt geometry). We study two ZnSe crystals, grown by the Bridgman method, whose linear and nonlinear optical properties are reported in Ref.~\cite{Mund20-ZnSe}. The samples provide two geometries where we can send the light either along the [001], or the [111] crystal axis. The magnetic field in the geometries is applied along the [100] and $[\bar{1}10]$ direction, respectively. In experiment, we either detect SHG intensity in dependence of emitted photon energy (spectrum), or in dependence of polarization of ingoing and outgoing photons (rotational anisotropy). Polarization dependences are measured in the parallel and crossed configurations for polarization angles of ingoing and outgoing photons $\psi(\varphi)=0^\circ,5^\circ,...,360^\circ$. In the first case, the polarizations are parallel ($\varphi=\psi$), i.e. $\textbf{E}^\omega\parallel\textbf{E}^{2\omega}$. In the latter case, the polarizations are perpendicular to each other ($\varphi=\psi+90^\circ$), $\textbf{E}^\omega\perp\textbf{E}^{2\omega}$. To test for a magnetic nonreciprocity, we measure the SHG signal for magnetic field in $\pm\textbf{x}$ direction, while keeping the direction of the light propagation the same. By this procedure the light ``sees'' the magnetic field from either the left, or the right side.

\section{Experimental results}

\begin{figure}[h]
	\begin{center}
		\includegraphics[width=0.48\textwidth]{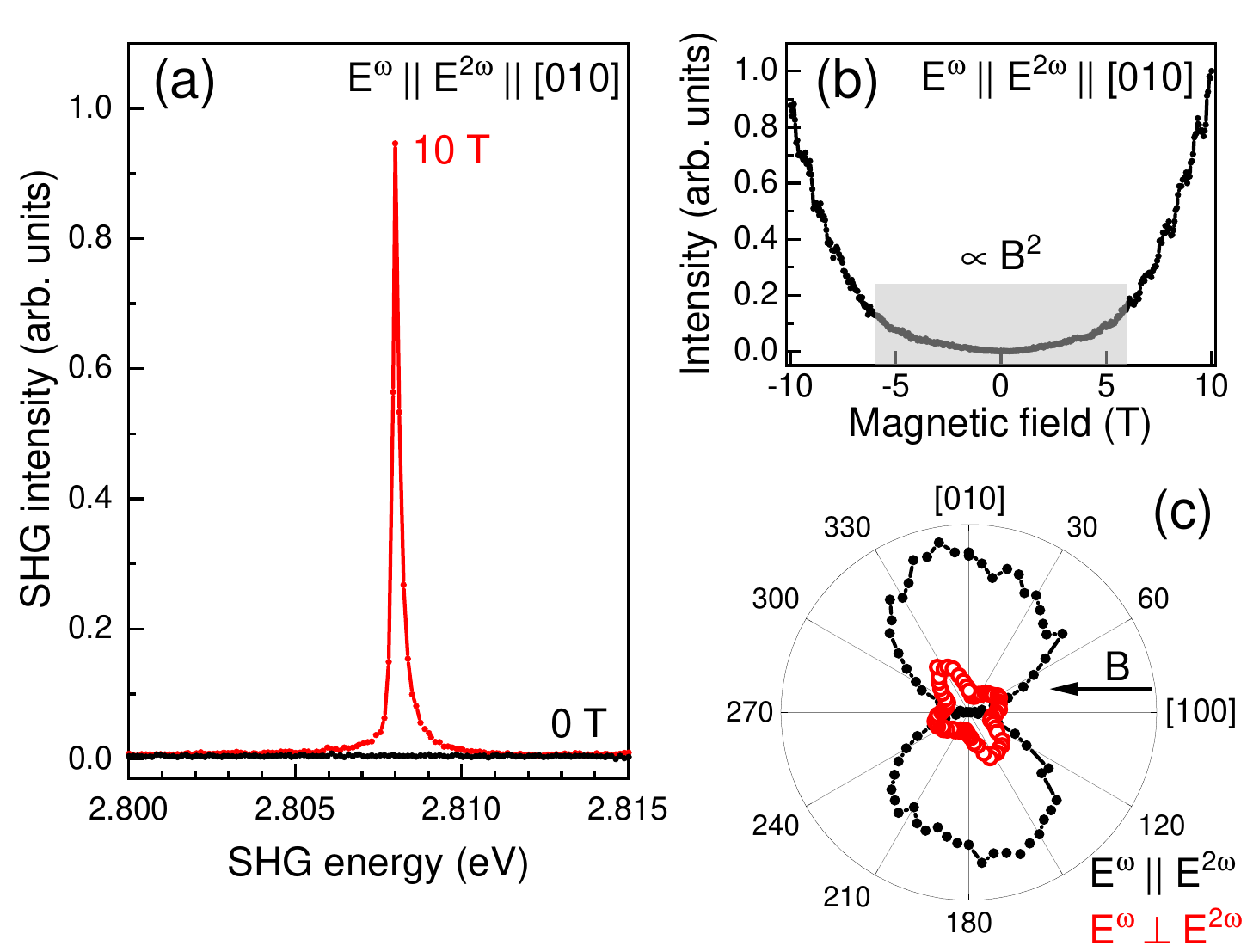}
		\caption{(a) SHG spectra with $\textbf{k}^\omega\parallel[001]$ at $B=0$~T (black) and magnetic field induced signal at $B=10$~T (red). (b) 1S SHG signal peak intensity normalized to its value at  $B=+10$~T in dependence on magnetic field strength. Signal in gray shaded area follows a B$^2$ dependence. (c) Rotational anisotropy of the 1S SHG resonance at $B=-10$~T. Closed black and open red dots are data for parallel $\textbf{E}^\omega\parallel\textbf{E}^{2\omega}$ and crossed $\textbf{E}^\omega\perp\textbf{E}^{2\omega}$ polarization configuration, respectively. $T=5$~K.}
		\label{pic.[001]_SHG_Bv-dep}
	\end{center}
\end{figure}

Figure~\ref{pic.[001]_SHG_Bv-dep}(a) shows SHG signal for the sample geometry $\textbf{k}^\omega\parallel\textbf{z}\parallel[001]$, $\textbf{B}\parallel\textbf{x}\parallel[100]$ and $\textbf{y}\parallel[010]$. At zero magnetic field no SHG signal is observed at the energy of the 1S exciton-polariton resonance. In fact, we do not observe it in the whole spectral range below and above the exciton resonance. This can be readily understood by the fact that, at $B=0$, the $\chi^{(2)}$ tensor, which describes the SHG process, has only one independent nonzero component, $\chi_{xyz}$ (the remaining nonzero components are obtained by permutations of its indices) \cite{Boyd, Birss}. Thus, at least one photon in the excitation, or emission process has to be polarized along the $z$-axis, which is not possible as the transversely polarized light is directed along that axis. The application of magnetic field induces SHG at the exciton resonance. At $B=10$~T a narrow line can be seen at $E_\textrm{1S}=2.80803$~eV for ingoing and outgoing light polarized along [010]. Its full width at half maximum (FWHM) is 0.21~meV. In ZnSe the SHG line is shifted by about 3 to 4~meV to higher energy from the exciton resonance in the reflectivity spectrum, which is due to the  exciton-polariton dispersion~\cite{Mund20-ZnSe}. Thus, the SHG involves the upper exciton-polariton branch. The magnetic field  dependence of the SHG intensity for the both field directions is shown in Fig.~\ref{pic.[001]_SHG_Bv-dep}(b). Whereas the intensity increases $\propto B^2$ in moderate fields up to about $\pm6$~T, it accelerates to $\propto B^4$ for higher fields. The magnetic-field-induced second-harmonic polarization, is described in analogy to Eq.~(\ref{eq.Pol_cryst}) by [cf. Refs.~\cite{Vollmer:2001aa,Kirilyuk:05,Kolmychek:2020aa}]
\begin{equation}\label{eq.Pol_mag}
P^{2\omega}_{i}(B)=\chi^{({eem})}_{ijkl}E^\omega_{j}E^\omega_{k}B_{l},
\end{equation} 
with two independent components $\chi^{({eem})}_{zzxx} = - \chi^{(3)}_{zzyy},$ and  $\chi^{({eem})}_{zxzx} = - \chi^{({eem})}_{zyzy}$ with the remaining ones obtained by the cyclic permutations of indices~\cite{Birss}. Since the SHG intensity is $\propto {|P^{2\omega}|^2}$, it increases as $B^2$. The $B^4$ increase of the SHG intensity hints at an additional SHG process, that is contributing particularly in stronger fields, or at the increased exciton oscillator strength. The analysis of the rotational anisotropies of the SHG in these geometries indicates that higher-order in $B$ contributions as well as possibly the contributions due to the radiation wavevector $\mathbf k^\omega$ can be important in this particular geometry, see SI for details.

Let us turn to the key experimental geometry with $\textbf{k}^\omega\parallel\textbf{z}^{\prime}\parallel[111]$, $\textbf{B}\parallel\textbf{x}^{\prime}\parallel[{1\bar{1}0}]$ and $\textbf{y}^{\prime}\parallel[11\bar{2}]$
, where SHG is symmetry allowed at zero magnetic fields. The photons are polarized in the $(x^{\prime}y^{\prime})$  plane. Thus, there is a projection of polarization to the crystal $[001]$ axis, and SHG is already observed in zero magnetic field due to $\chi_{xyz}$. 

\begin{figure}[h]
	\begin{center}
		\includegraphics[width=0.48\textwidth]{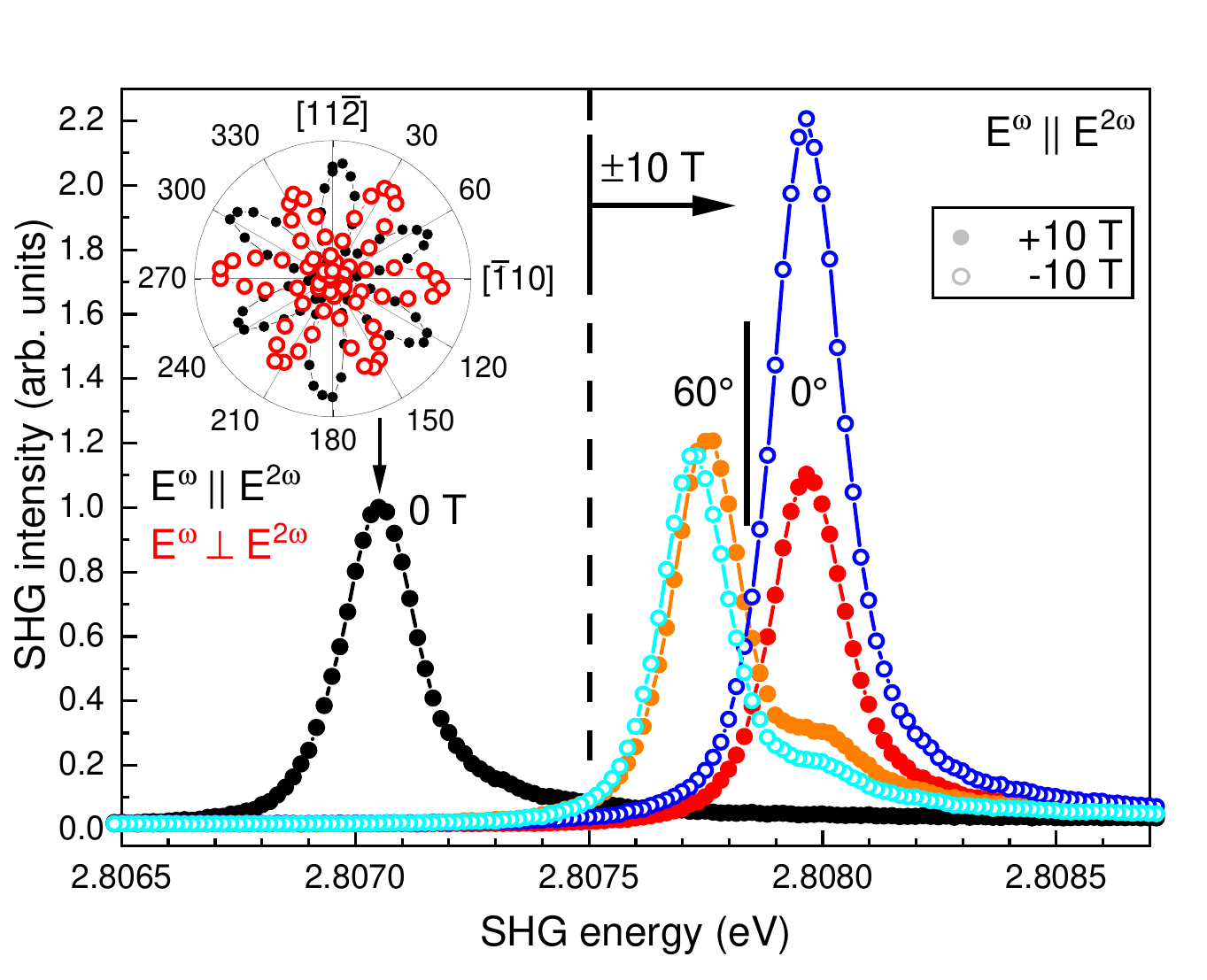}
		\caption{SHG spectra with $\textbf{k}^\omega\parallel[111]$ at $B=0$~T (closed black dots) and in $B=\pm10$~T. Spectra with closed dots correspond to $+10$~T (orange, red) and with open dots to $-10$~T (light blue, blue). All spectra are detected in parallel configuration, $\textbf{E}^\omega\parallel\textbf{E}^{2\omega}$. Lines at $B=0$~T and higher energetic line in magnetic field are detected with $\varphi=\psi=0^\circ$ and lower energetic line in magnetic field with $\varphi=\psi=60^\circ$. The inset shows the rotational anisotropy of the 1S resonance at $B=0$~T. Closed black and open red dots give data for parallel $\textbf{E}^\omega\parallel\textbf{E}^{2\omega}$ and crossed $\textbf{E}^\omega\perp\textbf{E}^{2\omega}$ polarization configuration, respectively.}
		\label{pic.[111]_SHG_specs}
	\end{center}
\end{figure}

In Figure~\ref{pic.[111]_SHG_specs}, we show the SHG signal at the 1S exciton resonance at $B=0$~T (black) at $E_\textrm{0T}=2.80705$~eV with a FWHM of 0.18~meV that is slightly lower in energy than in the [001] geometry. In comparison with the resonant SHG at the 1S resonance, the non-resonant SHG has much lower intensity which appears as zero (noise) level in the spectra. It is compared with signal in $B=+10$~T (red, orange) and $B=-10$~T (blue, light blue). The six-fold rotational anisotropy of the 1S resonance in zero magnetic field shown in the inset of Fig.~\ref{pic.[111]_SHG_specs}  is perfectly reproduced by a simulation taking into account the crystallographic $\chi_{xyz}$ component, see Appendix~\ref{sec:model}. This confirms that the SHG excitation and emission are due to virtual transitions allowed in the electric-dipole approximation. 

At high magnetic field $B=\pm10$~T, we observe two resonances split by 0.23~meV at $E_{60^\circ}=2.80773$~eV and $E_{0^\circ}=2.80797$~eV with comparable FWHM as at 0~T. They are related to the exciton fine structure: In the absence of magnetic field the optically active $1S$ exciton is three-fold degenerate. The polariton effects (in our case $\mathbf k\parallel \mathbf z'$ splits the $\mathbf z$-polarized states (longitudinal state) from the degenerate doublet $\mathbf x',\mathbf y'$.  The application of the magnetic field splits the states according to their spin angular momentum component along the field, $\mathbf x^\prime$ axis. The state polarized along the $\mathbf x^\prime$ experiences only a weak diamagnetic shift. The $\mathbf z^\prime$ and $\mathbf y^\prime$ polarized states become mixed. Hence, we assign the lower energy resonance to the $\textbf{x}^{\prime}$, i.e. [$\bar{1}$10], component of the 1S exciton, Fig.~\ref{pic.[111]_SHG_specs}. Naturally, this contribution is polarized parallel to the magnetic field and experiences relatively weak diamagnetic shift. The higher energy resonance corresponds to the $\textbf{y}^{\prime}$, i.e. [11$\bar{2}$], component of the 1S exciton, which is getting mixed with the $\textbf{z}^{\prime}$ component by the Zeeman effect. It is polarized perpendicular to the magnetic field and, therefore, shifted by the Zeeman effect in addition to the diamagnetic shift. As already stated, both components are shifted to higher energy by the exciton-polariton dispersion. The $\textbf{z}^{\prime}$ component (with a small admixture of $\textbf{y}^{\prime}$), however, is not observed because it is mainly the longitudinal state with respect to the light propagation direction. We expect it at about 2.803~eV with low intensity due to only a small admixture of $\textbf{y}^{\prime}$, see the model below. The $\textbf{x}^{\prime}$ and $\textbf{y}^{\prime}$ components can be very well separated by choosing the angles of polarization for ingoing and outgoing photons. Their parallel rotational anisotropies are shown in Figs.~\ref{pic.[111]_SHG_Bv-dep}(c) and \ref{pic.[111]_SHG_Bv-dep}(d), respectively, confirmed by the model, see below. Thus, the low energy $\textbf{x}^{\prime}$ exciton component appears for $\textbf{E}^\omega\parallel\textbf{E}^{2\omega}\parallel60^\circ$ and $\textbf{y}^{\prime}$ has the strongest intensity for $\textbf{E}^\omega\parallel\textbf{E}^{2\omega}\parallel0^\circ$.

Importantly, the intensities of the SHG components are strongly nonreciprocal functions of the magnetic field. The main finding of our study is that the $\textbf{y}^{\prime}$ exciton component is only slightly increased in intensity for positive magnetic field, whereas its intensity is more than doubled in negative field, see Figs.~\ref{pic.[111]_SHG_specs} and \ref{pic.[111]_SHG_Bv-dep}(b). In contrast, the $\textbf{x}^{\prime}$ component shows almost no intensity change for either direction of the magnetic field, see Fig.~\ref{pic.[111]_SHG_Bv-dep}(a).

\begin{figure}[h]
	\begin{center}
		\includegraphics[width=0.48\textwidth]{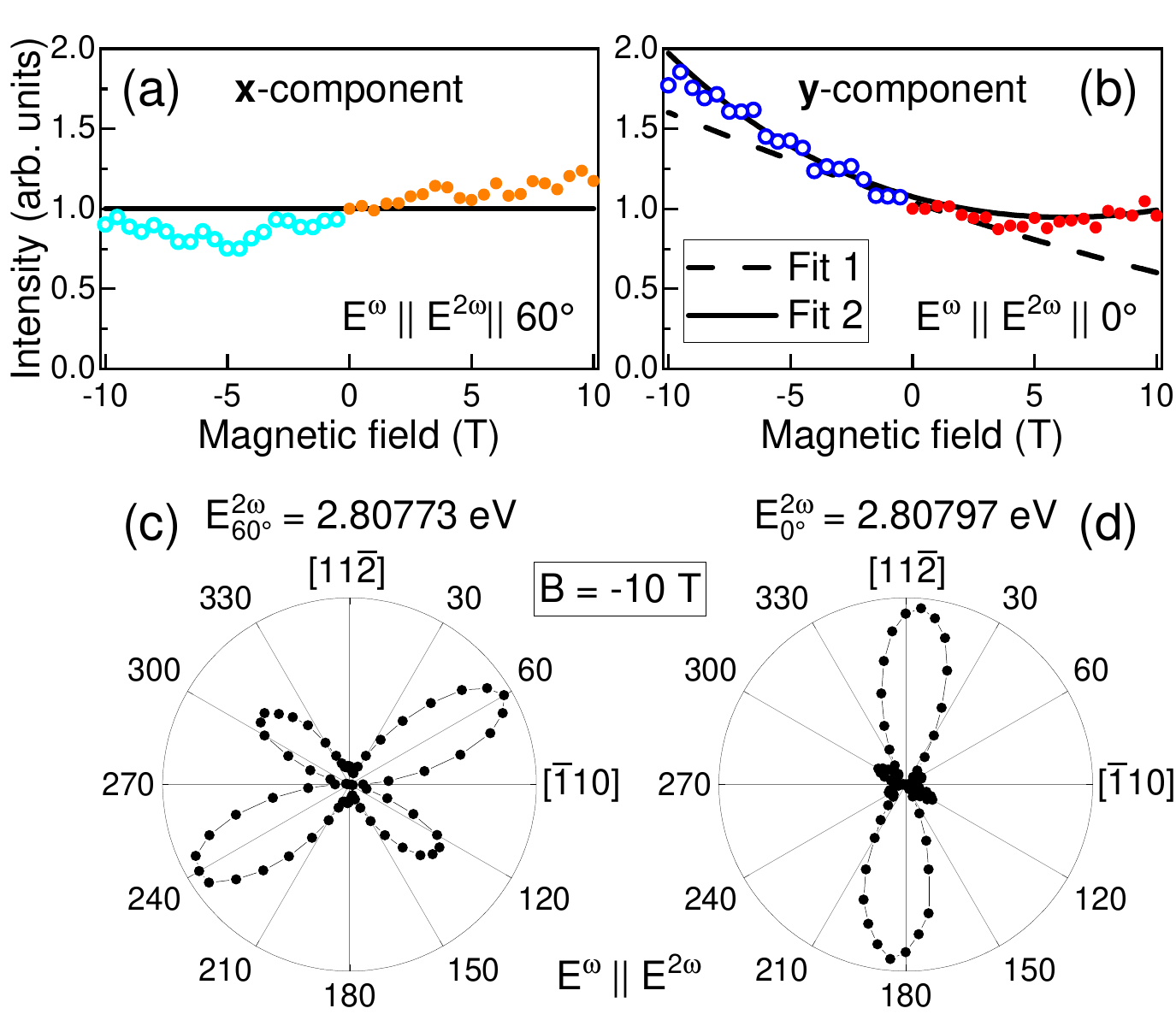}
		\caption{SHG 1S resonance peak intensity in dependence of magnetic field for (a) $\textbf{E}^\omega\parallel\textbf{E}^{2\omega}\parallel60^\circ$ and (b) $\textbf{E}^\omega\parallel\textbf{E}^{2\omega}\parallel0^\circ$. The solid black line in (a) gives intensity level at $B=0$~T. 
        "Fit 1" and "2" in (b) are fitting functions to the data by Eq.~(\ref{Fit:2}). Parallel rotational anisotropies of the 1S at $B=-10$~T at (c) $E^{2\omega}=2.80773$~eV and (d) $E^{2\omega}=2.80797$~eV. Closed black dots give data for parallel $\textbf{E}^\omega\parallel\textbf{E}^{2\omega}$ polarization configuration. 
        }
		\label{pic.[111]_SHG_Bv-dep}
	\end{center}
\end{figure}


The dependence of the 1S resonance intensity on the magnetic field up to $\pm10$~T is shown for the $\textbf{x}^{\prime}$ and $\textbf{y}^{\prime}$ components in Figs.~\ref{pic.[111]_SHG_Bv-dep}(a) and \ref{pic.[111]_SHG_Bv-dep}(b), respectively. The intensity of the $\textbf{x}^{\prime}$ component changes by about $25~\%$ up to $\pm10$~T. In contrast, we observe an increase by $100~\%$ at $-10$~T and no change in intensity at $+10$~T for the $\textbf{y}^{\prime}$ component. The respective rotational anisotropies in Figs.~\ref{pic.[111]_SHG_Bv-dep}(c) and \ref{pic.[111]_SHG_Bv-dep}(d) show that the components, which are split in energy by the magnetic field, give in sum the unperturbed anisotropy at $B=0$~T. More precisely, at $B=0$~T both states are degenerate and their emission interferes, which results in the six-fold anisotropy shown in the  insert of Fig.~\ref{pic.[111]_SHG_specs}. The magnetic field lifts their degeneracy and the individual states anisotropies can be seen.

\section{Discussion}

We start with the phenomenological analysis based on the symmetry arguments to describe the non-reciprocal SHG observed in experiments. In the axes frame $\mathbf x||[1\bar10]$, $\mathbf y||[11\bar2]$ and $\mathbf z\parallel [111]$ (hereafter the primes are omitted for the sake of brevity) the polarization induced in the second order in the electric field and the first order in the static magnetic field reads
\begin{align}
&P_x^{2\omega} ={ (\chi_0+\chi_2 B_x)  2} E_x^\omega E_y^\omega, \label{P:111}
 \\
&P_y^{2\omega} = (\chi_0+\chi_2 B_x)({E_x^{\omega}}^2-{E_y^{\omega}}^2) + \chi_1B_x ({E_x^{\omega}}^2+{E_y^{\omega}}^2).\nonumber
\end{align}
The parameter $\chi_0$ describes the crystallographic SHG generation, Eq.~\eqref{eq.Pol_cryst}, and two parameters $\chi_1$ and $\chi_2$  describe the linear-in-$B$ contributions, Eq.~\eqref{eq.Pol_mag}.  In Eqs.~\eqref{P:111} we have included only $x$- and $y$-components of the polarization because only those components contribute to the measured signal due to the transversality of the electric field. It immediately follows from Eqs.~\eqref{P:111} that the polarization at a double frequency contains $B$-independent and $B$-linear contributions, thus the intensity of SHG detected in the experiment can be recast as {$I = |\mathcal{A} + \mathcal{C} B|^2=\mathcal |A|^2 + \mathcal |C|^2 B^2 + 2 {\rm Re}({\mathcal A}^* \mathcal C) B$} and contains $B$-independent, $B^2$, and, importantly, the  $B$-linear interference-contribution with constants $\mathcal{A} \propto \chi_0$ and $\mathcal{C}$ being proportional to the combination of $\chi_1$ and $\chi_2$. Such dependence is also typical for the nonlinear magneto-optical Kerr effect~\cite{Murzina:1998aa,Kolmychek:2020aa}. The corresponding fit is shown by a dashed line (Fit 1) in Fig.~\ref{pic.[111]_SHG_Bv-dep}(b). The analysis of the polarization dependence and comparison of Fig.~\ref{pic.[111]_SHG_Bv-dep}(a) and \ref{pic.[111]_SHG_Bv-dep}(b) demonstrates that $\chi_2$ is negligible because the nonreciprocal effect in $\mathbf x$ polarization is almost absent, while the ratio $\chi_1/\chi_0 \approx -0.024/{\mathrm T}$. It is in agreement with the model analysis in Appendix~\ref{sec:model} where $\chi_2=0$.
The Fit-1 already illustrate the non-reciprocal dependence of the SHG on the magnetic field, but deviates from the experimental data at relatively high magnetic fields of several Teslas. The accuracy of fitting can be improved if one includes the $B^2$ contributions to the susceptibility and presenting the intensity as
\begin{equation}
\label{Fit:2}
I = {|}\mathcal{A} + \mathcal{C} B + \mathcal{C}' B^2{|}^2.
\end{equation}
Here the term with parameter $\mathcal{C}'$ can be related to the diamagnetic effect and also independent $B^2$ contributions to the nonlinear susceptibility. The curve Fit 2 in Fig.~\ref{pic.[111]_SHG_Bv-dep}(b) is calculated after Eq.~\eqref{Fit:2} with $\mathcal C'/\mathcal{A} \approx 7.9{\times 10^{-4}/{\mathrm T}^2}$, and $\mathcal{C}/\mathcal{A} =0.02/\mathrm T$.

\begin{figure}[h]
	\begin{center}
		\includegraphics[width=0.48\textwidth]{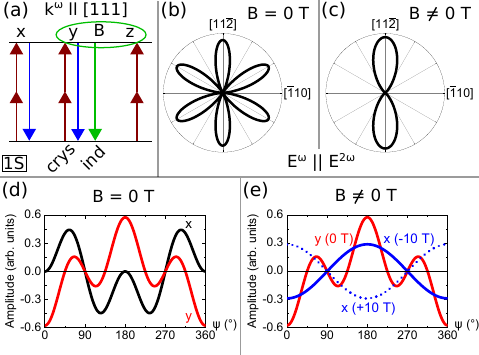}
		\caption{(a) Two-photon excitation and one-photon emission paths for the three 1S-components. Magnetic field mixes y- and z-component and allows for induced SHG emission. SHG rotational anisotropy for parallel configuration ($\textbf{E}^\omega\parallel\textbf{E}^{2\omega}$) for (b) crystallographic and (c) magnetic field induced contribution. 
        Phases of SHG with light polarization detection parallel to excitation polarization from x- and y-components (d) and of crystallographic and induced SHG from y-component (e). Blue dotted and solid line give phases for positive and negative field, respectively}
		\label{pic.interference_scheme}
	\end{center}
\end{figure}

The microscopic mechanism responsible for the non-reciprocal effect is illustrated in Fig.~\ref{pic.interference_scheme}(a) and related to the Zeeman effect of the magnetic field. The triplet of optically active $1S$ excitons at $B=0$ is formed from the states $|x\rangle$, $|y\rangle$ and $|z\rangle$ one-photon dipole active in the corresponding linear polarization. These states are also two-photon active according two $|x\rangle \propto E_x E_y$, $|y\rangle \propto E_x^2 - E_y^2$, and $|z\rangle \propto E_x^2+E_y^2$, proving the contribution $\chi_0$ in the phenomenological relation of Eq.~\eqref{P:111}, see corresponding polarization dependence in Figs.~\ref{pic.interference_scheme}(b,d). In the bulk crystal the nonlinear optical response is provided by the exciton-polaritons~\cite{Warkentin18,Mund20-ZnSe}. The analysis of the polariton dispersion relation and phase-matching conditions demonstrates that the upper polariton branch is involved in the SHG and, correspondingly, the longitudinal $z$-state is below the $x$, $y$-polarized ones. Hence, at $\mathbf B\parallel \mathbf x$ the optically active $|y\rangle$ exciton-polariton state acquires an admixture of the $|z\rangle$,
\begin{equation}
\label{Zeeman}
|\tilde y\rangle \approx |y\rangle + c_{yz} |z\rangle,
\end{equation}
with the following estimation for the admixture coefficient, see see Appendix~\ref{sec:model} for details,
\begin{equation}
\label{Zeeman2}
c_{yz} \approx {\mathrm i} \frac{g\mu_B B_x}{\Delta} .
\end{equation}
where $\mu_B$ is the Bohr magneton, $g$ is the exciton Land\'e factor and $\Delta$ is the frequency-dependent splitting between $y$- and $z$-polarized states due to the polariton effects ($\Delta \approx 5$~meV, see above). Additional contribution to $c_{yz}$ results from the magneto-spatial dispersion contribution, see Appendix~\ref{sec:model} for details.

This mixing provides $\chi_1 \ne 0$ in Eq.~\eqref{P:111}, compare Figs.~\ref{pic.[111]_SHG_Bv-dep}(d) and \ref{pic.interference_scheme}(c) and, correspondingly, a nonreciprocal dependence of the susceptibility on the magnetic field observed in the experiment, see also Fig.~\ref{pic.interference_scheme}(e).  Note that the $|z\rangle$ state acquires $B$-linear admixture of the $|y\rangle$ state, cf. Eq.~\eqref{Zeeman}, but this state does not take part in the SHG at $B=0$ because it is polarized along the wave vector. Thus, its contribution to the SHG intensity is quadratic in $B$. At the same time, $\chi_2 =0$ in this mechanism, because to the linear order in $B$ the $|x\rangle$ state is unaffected by the Zeeman effect and, accordingly, the $B_y$-linear contribution to $P_{x}^{2\omega}$ does not appear. This allows us to conclude, that the interference of crystallographic and magnetic field induced contributions is the origin of the measured SHG nonreciprocity. The Zeeman effect is an origin for the magnetic field induced SHG in our case.


Since the breaking of reciprocity is connected to the 1S exciton one would expect to observe nonreciprocity up to the temperature where the 1S is dissociated. From its binding energy of 20~meV we estimate the temperature to about 200~K. This is much higher temperature than for the recently reported SHG nonreciprocity in CuB$_2$O$_4$ originated from the interference of crystallographic and induced toroidal dipole moment SHG~\cite{Mund20-CuB4O2, Toyoda20}. In this materials the SHG nonreciprocity is not detected at temperatures exceeding the Ne\'el temperature of 20~K, i.e. in the paramagnetic phase of CuB$_2$O$_4$.

\section{Conclusion}

We found experimentally a nonreciprocal magnetic field induced second harmonic generation at the 1S exciton-polariton in ZnSe crystals with the $\textbf{k}^\omega\parallel[111]$ orientation. This is achieved by applying a magnetic field perpendicular to the light $\textbf{k}^\omega$ in either direction, which is comparable to a reversal of the light direction through an unchanged system. Interference of crystallographic and magnetic-field-induced SHG signals provides different SHG intensities for opposite field directions for certain polarization angles manifesting nonreciprocity as the nonlinear magneto-optical Kerr effect. As a confirmation of the mechanism, we choose the geometry with $\textbf{k}^\omega\parallel[001]$ where crystallographic SHG is forbidden. Accordingly, no nonreciprocity is observed in this geometry.   Thus, we confirm that a single contribution to SHG, namely the magnetic field induced one, is not sufficient to observe nonreciprocity. Such interference of the  crystallographic and magnetic-field-induced SHG provides a tool for studying the symmetry and mixing of exciton-polariton states. To the best of our knowledge this is the first experimental observation of the nonreciprocal SHG in semiconductor crystals, and the first one for exciton-polaritons.

We are firm, that the nonreciprocal SHG can be observed in a manifold of semiconductor crystals, especially with pronounced exciton-polaritons. One can expect that in systems with significantly higher exciton binding energies such as ZnO and Cu$_2$O, where the thermal dissociation of excitons is prevented, the effect can be observed up to room temperature.

\textbf{ORCID}\\

Johannes Mund:      0000-0002-8022-7584  \\  
Dmitri R. Yakovlev: 0000-0001-7349-2745 \\  
Andreas Farenbruch:	0000-0001-9863-8755 \\  
Nikita Siverin:		0000-0002-4643-845X \\  
Marina A. Semina:   0000-0003-3796-2329    \\ 
Mikhail M. Glazov:  0000-0003-4462-0749    \\ 
E. L. Ivchenko:     0000-0001-7414-462X   \\  
Manfred~Bayer:   0000-0002-0893-5949 \\ 

{\bf Acknowledgments.}
J.M., D.R.Y, A.F., N.V.S., and M.B. acknowledge the financial support by the Deutsche Forschungsgemeinschaft in the frame of the Collaboration Research Center TRR142 (project A11). Theoretical contribution of M.A.S, M.M.G., and E.L.I. was supported by RSF Project 23-12-00142.


\appendix

\section{Microscopic model}\label{sec:model}

The effective Hamiltonian of the $1S$ exciton-polariton in the presence of magnetic field $\mathbf B$ can be written as 
\begin{equation}
    \label{Ham}
    \mathcal H = \Delta L_z^2 + g\mu_B \mathbf B \cdot \mathbf L,
\end{equation}
where $\Delta$ is the constant describing the splitting between the $x,y$ and $z$ polarized exciton-polaritons (wavevector $\mathbf k\parallel \mathbf z$), $g$ is the exciton Land\'e factor, and $\mu_B$ is the Bohr magneton. The matrices 
\begin{equation} \label{I}
L_x =  \left[ \begin{array}{ccc} 0&0&0\\ 0&0&-{\rm i}\\ 0&{\rm i}&0 \end{array} \right]\:,\:
 L_y =  \left[ \begin{array}{ccc} 0&0&{\rm i}\\ 0&0&0\\ -{\rm i} &0&0 \end{array} \right]\:,\: L_z =  \left[ \begin{array}{ccc} 0&-{\rm i}&0\\ {\rm i}&0&0\\ 0&0&0 \end{array} \right]\: \,,
 \end{equation}
 are the angular momentum-1 matrices in the Cartesian basis (the order of the basic states is $x$, $y$ and $z$). The terms responsible for the diamagnetic shift are omitted. In the $B$-linear regime the $x$-polarized exciton state is not perturbed by the field, while the $y$- and $z$- polarized states become intermixed. Particularly, the perturbed $|\tilde y\rangle$ state reads
 \begin{equation}
\label{Zeeman}
|\tilde y\rangle \approx |y\rangle + {\mathrm i}\frac{g\mu_B B_x}{\Delta} |z\rangle,
\end{equation}
in agreement with Eq. (5) of the main text. The splitting between the $|\tilde y\rangle$ and non-perturbed $|\tilde x\rangle = |\tilde x\rangle$ states is given by $(g\mu_B B)^2/\Delta$.

In what follows we use the coordinate frame  $\mathbf x \parallel [1\bar10]$, $\mathbf y \parallel [11\bar2]$ and $\mathbf z \parallel [111]$ with $\mathbf B\parallel \mathbf x$ where the nonreciprocity is experimentally observed. The second harmonic generation effect is the two-photon excitation of a given state followed by its coherent single-photon emission. The selection rules for the two-photon transitions are the follows: the $|y\rangle$ is excited by the field combination $E_x^2 - E_y^2$, while the $|z\rangle$ state is excited by the combination $E_x^2+E_y^2$. In order to calculate the second-order susceptibility we recast the wavefunction of the crystal in the form
\begin{equation}
    \label{psi:cr}
    \Psi(t) = |0\rangle + (C_x |x\rangle + C_y |y\rangle + C_z |z\rangle)e^{-2\mathrm i \omega t},
\end{equation}
where $|0\rangle$ is the ground state, $C_\alpha$ ($\alpha=x,y,z$) are the coefficients, $|C_\alpha|\ll 1$. Considering the two-photon transitions with intermediate states in the remote bands we derive the following equations for the coefficients
\begin{equation}
    \label{eq:C}
    2\hbar\omega \hat C = \mathcal H \hat C + \mathcal V^{(2)},
\end{equation}
where $\hat C=[C_x,C_y,C_z]^T$ and
\begin{equation}
    \label{V2}
    \mathcal V^{(2)} = M_{1S}^{(2)}
    \begin{pmatrix}
    2E_x^\omega E_y^\omega & 0 & 0\\
    0 & {E_x^{\omega}}^2-{E_y^{\omega}}^2 & 0\\
    0 & 0 & {E_x^{\omega}}^2+{E_y^{\omega}}^2
    \end{pmatrix},
\end{equation}
with $M_{1S}^{(2)}$ being the two-photon excitation matrix element of the $1S$ exciton and $E_{x,y} \equiv E_{x,y}^\omega$ being the field components of the fundamental frequency radiation inside the crystal, describes the two-photon driving. Solving Eq.~\eqref{eq:C} for the coefficients $C_\alpha$ in the linear-in-$B$ regime and evaluating the $x$- and $y$-components of the excitonic polarization we arrive at
\begin{subequations}
        \label{Polarization}
        \begin{eqnarray}
               P_x^{2\omega}& =& \frac{d_{1S} M^{(2)}_{1S}}{2\hbar\omega - \mathcal E_{x} + \mathrm i \Gamma}\times 2  E_x^\omega E_y^\omega,\\
            P_y^{2\omega}& =& \frac{d_{1S} M^{(2)}_{1S}}{2\hbar\omega - \mathcal E_{y} + \mathrm i \Gamma} \left[{E_x^{\omega}}^2-{E_y^{\omega}}^2\right. \label{Py}\\
            &&\left.- \frac{\mathrm i g\mu_B B}{2\hbar\omega - \mathcal E_{z}+ \mathrm i \Gamma}\left({E_x^{\omega}}^2+{E_y^{\omega}}^2 \right)\right] \,.\nonumber
               \end{eqnarray} 
\end{subequations}
Here $\mathcal E_{\alpha}$ and $\Gamma$ are the exciton energy and damping, respectively, and $d_{1S}$ is the microscopic dipole moment of the $1S$ exciton. In the notations of Eq.~(3) of the main text 
\begin{subequations}
    \label{suscpts}
    \begin{align}
        \chi_0 = \frac{d_{1S}M_{1S}^{(2)}}{2\hbar\omega - \mathcal E_{y} + \mathrm i \Gamma} , \\
        \chi_1 = - \frac{\mathrm i g\mu_B}{2\hbar\omega - \mathcal E_{z}+ \mathrm i \Gamma} \chi_0, \label{suscpts:1} \\
        \chi_2 =0.
    \end{align}
\end{subequations}
In the absence of damping, $\Gamma=0$, the phases of $\chi_0$ and $\chi_1$ are different by $\pm \pi/2$, note the factor $\mathrm i$ in Eqs.~\eqref{Zeeman} and \eqref{suscpts:1}. It means that  while the $B$-linear contribution is present in the second harmonic polarization, it vanishes in the second harmonic intensity. In this case the resulting SHG is reciprocal. Inclusion of nonzero $\Gamma$ enables the interference of the $B$-independent and $B$-linear terms in Eq.~\eqref{Py} giving rise to the nonreciprocal SHG.

To illustrate the effect let us calculate the SHG intensity in the $y$-polarization at $x$-polarized incident radiation:
\begin{multline}
    \label{SHG:I}
    I^{2\omega}\propto |P^{2\omega}_y|^2 =|\chi_0 E_x^\omega|^2|1+\chi_1B|^2 \\
    \propto \frac{1}{(2\hbar\omega - \mathcal E_y)^2+\Gamma^2}\left|1- \frac{\mathrm i g \mu_B B}{2\hbar\omega - \mathcal E_z + \mathrm i \Gamma} \right|^2  \,.
\end{multline}
In the $B$-linear approximation we obtain
\begin{multline}
    \label{SHG:I:1}
    I^{2\omega} 
    \propto \frac{1}{(2\hbar\omega - \mathcal E_y)^2+\Gamma^2}\left(1+ \frac{2 g \mu_B B\Gamma}{(2\hbar\omega - \mathcal E_z)^2+ \Gamma^2} \right).
\end{multline}
This expression can be further simplified for the resonant case where $|2\hbar\omega -\mathcal E_y|\ll \Delta$ with the result
\begin{equation}
    \label{SHG:I:2}
    I^{2\omega}\propto 
     \frac{1}{(2\hbar\omega - \mathcal E_y)^2+\Gamma^2}\left(1+ \frac{2 g \mu_B B\Gamma}{\Delta^2+ \Gamma^2} \right).
\end{equation}
Equations~\eqref{SHG:I:1} and \eqref{SHG:I:2} clearly show that the nonreciprocal behavior of the SHG is possible provided that the damping $\Gamma$ of excitons is taken into account. In this way, the expressions for the intensity are time-reversal invariant.

Additional contributions to the nonreciprocal in $\mathbf B$ SHG can result from the $\mathbf k\cdot \mathbf B$ linear terms in the exciton Zeeman Hamiltonian similarly to the contributions responsible for the magneto-spatial dispersion~\cite{gogolin_magn_eng,1998JETP...87..553K,PhysRevB.88.155203}. Particularly, for $\mathbf k\parallel z$ and $\mathbf B \parallel x$ there are two corresponding contributions to the exciton effective Hamiltonian (in our reference frame):
\begin{equation}
    \label{H:kB}
    \mathcal H_{kB} = \gamma_1 k_z B_x\frac{L_yL_z+L_zL_y}{2} + \gamma_2 k_z B_x(L_x^2 - L_y^2),
\end{equation}
where the matrices $L_x,L_y,L_z$ are given by Eq.~\eqref{I} and $\gamma_{1,2}$ are the coefficients. The term $\propto \gamma_2$ contributes to the splitting of $x$- and $y$-polarized excitons, while the term $\gamma_1$ induces additional mixing of the $y$- and $z$-polarized states as
 \begin{equation}
\label{Zeeman:1}
|\tilde y\rangle \approx |y\rangle + \left({\mathrm i}\frac{g\mu_B B_x}{\Delta} - \frac{\gamma_1 k_z B_x}{2\Delta}\right) |z\rangle,
\end{equation}
In contrast to the Zeeman effect this term does not contain the imaginary unity. It yields additional contribution to $\chi_1$, Eq.~\eqref{suscpts:1}, in the form
\begin{equation}
            \chi_1' = \frac{1}{2} \frac{\gamma_1 k}{2\hbar\omega - \mathcal E_{z}+ \mathrm i \Gamma} \chi_0, \label{suscpts:1:1}
\end{equation}
with the same phase as $\chi_0$. These terms interfere in SHG providing the nonreciprocal $k_z B_x$ contribution in the form ($|2\hbar\omega - \mathcal E_y|\ll \Delta$)
\begin{equation}
    \label{SHG:I:2:1}
    I^{2\omega'}\propto 
     \frac{1}{(2\hbar\omega - \mathcal E_y)^2+\Gamma^2}\frac{\gamma_1 {k_z B_x}\Delta }{\Delta^2+\Gamma^2}.
\end{equation}

\end{document}